\documentclass[12pt,a4paper,prd,superscriptaddress,nofootinbib]{revtex4}
\pdfoutput=1
\usepackage[latin1]{inputenc}
\usepackage[english]{babel}
\usepackage{amssymb}
\usepackage{amsmath}
\usepackage{amsthm}
\usepackage[]{graphicx}
\usepackage[]{subfigure}
\graphicspath{{./figs/}}
\usepackage{tensor}
\usepackage{color}
\usepackage{cancel}
\usepackage{setspace}
\usepackage{fancyhdr}
\usepackage[bookmarks,linktocpage,colorlinks=true,plainpages=false,citecolor=blue,linkcolor=blue,urlcolor=blue,filecolor=blue]{hyperref}

\newcommand{\bea}{\begin{eqnarray}}
\newcommand{\eea}{\end{eqnarray}}
\newcommand{\be}{\begin{equation}}
\newcommand{\ee}{\end{equation}}
\newcommand{\beast}{\begin{eqnarray*}}
\newcommand{\eeast}{\end{eqnarray*}}
\newcommand{\pkt}{\; .}
\newcommand{\kma}{\; ,}
\newcommand{\nn}{\nonumber}
\def\e{{\rm e}}

\begin{document}

\title{Towards a Solution of the Negative Mode Problem\\
       in Quantum Tunnelling with Gravity}

\author{Michael Koehn}
\email{koehn@physics.upenn.edu}
\affiliation{Center for Particle Cosmology, Department of Physics \& Astronomy, University of Pennsylvania,
Philadelphia, PA 19104-6395, U.S.A.}

\author{George Lavrelashvili}
\email{lavrela@itp.unibe.ch }
\affiliation{Department of Theoretical Physics, A.\,Razmadze Mathematical Institute, I.\,Javakhishvili Tbilisi State University,
GE-0177 Tbilisi, Georgia}
\affiliation{Max Planck Institute for Gravitational Physics, Albert Einstein Institute, D-14476 Potsdam, Germany}

\author{Jean-Luc Lehners}
\email{jean-luc.lehners@aei.mpg.de}
\affiliation{Max Planck Institute for Gravitational Physics, Albert Einstein Institute, D-14476 Potsdam, Germany}

\date{\today}

\begin{abstract}
\vspace{1cm}
In the absence of gravity, one can prove that tunnelling instantons exhibit exactly one negative mode in their spectrum of fluctuations. It is precisely the existence of this tunnelling negative mode that warrants an interpretation of these solutions as mediating the decay of a metastable vacuum. In the presence of gravity the situation is much more subtle, not least because of diffeomorphism invariance. New complications arise here: in particular, the kinetic term of the fluctuations can change sign somewhere along the instanton. We show that in this case the mode functions remain non-singular, and the tunnelling negative mode continues to exist. Moreover, the eigenvalues vary continuously when the potential is varied such that the kinetic term of the fluctuations switches sign. However, the negative kinetic term implies the additional existence of an infinite tower of negative modes, whose significance and interpretation remain elusive.
\end{abstract}

\maketitle

\newpage

\section{Introduction}

Since the pioneering work of Coleman and De Luccia (CdL) \cite{Coleman:1980aw} many articles were devoted to the investigation of
metastable vacuum decay with gravity.
The recent discovery of a Higgs boson at the LHC highlights the importance of knowing all possible
corrections (including gravitational ones) to the vacuum decay rate, given that the present Higgs boson and top quark mass values
indicate that we may live in a metastable vacuum
\cite{Isidori:2001bm,Isidori:2007vm,EliasMiro:2011aa,Degrassi:2012ry,Kobakhidze:2013tn,Branchina:2013jra,Kobakhidze:2014xda,Branchina:2014usa,Branchina:2014rva,Loebbert:2015eea,Bezrukov:2012sa,Hamada:2014iga,Bezrukov:2014bra}.   In spite of many investigations of various aspects of metastable vacuum decay with gravity\footnote{Aspects of interest, but not dealt with in detail here, include \cite{Lee:2008hz,Lee:2009bp,Lee:2011ms,Battarra:2014naa}.}, there remain several important open questions in the field. One of them is the negative mode problem.
While in flat space-time it is relatively easy to show that a bounce solution describing metastable vacuum decay \cite{Coleman:1977py,Callan:1977pt,Coleman:1978ae} contains exactly one eigenmode with negative eigenvalue (we refer to this as the \emph{tunnelling negative mode})
in its spectrum of linear perturbations (as it should - see \cite{Coleman:1987rm}), once gravity is included this question is still not completely solved
\cite{Lavrelashvili:1985vn,Tanaka:1992zw,Lavrelashvili:1998dt,Tanaka:1999pj,Khvedelidze:2000cp,Lavrelashvili:1999sr,Gratton:2000fj,Hackworth:2004xb,
Lavrelashvili:2006cv,Dunne:2006bt,Brown:2007sd,Yang:2012cu,Battarra:2013rba,Lee:2014uza,Lee:2014_PhD}.

The quadratic action for small perturbations about CdL bounces was first calculated by Lavrelashvili, Rubakov and Tinyakov (LRT) \cite{Lavrelashvili:1985vn} using a Lagrangian formulation and fixing the gauge in order to obtain an action for the single remaining degree of freedom. It was found that this action has the following structure:
one obtains a factor $Q_{LRT}$, given by Eq.~(\ref{eq:Qlrt}) below, in front of the kinetic term of the perturbations. This factor $Q_{LRT}$
depends on a certain combination of background quantities and typically becomes negative on some interval along the bounce trajectories.
This specific gauge choice was criticised in \cite{Tanaka:1992zw}, where it was argued that the gauge is ill-defined at the middle of the instanton, i.e. at the location of maximal 3-volume of the geometry.\footnote{This is reminiscent of the issue of gauge fixing in bouncing universes, where the scale factor also passes through an extremum - see e.g. \cite{Battarra:2014tga}.} In fact, in the Lagrangian approach, where only configuration variables are involved, many gauge choices lead to similar kinds of trouble. Motivated by this, Tanaka and Sasaki (TS) \cite{Tanaka:1992zw} derived a quadratic action using the
Hamiltonian formalism. After excluding matter degrees of freedom TS obtained the action for one
(gravitational) degree of freedom and showed that there are no negative modes associated with this quadratic action
\cite{Tanaka:1992zw,Tanaka:1999pj}.
Realizing that the approach of TS is not fully satisfactory, in particular because it does not allow one to recover the flat space limit when gravity is turned off and because it hinges on a number of delicate analytic continuations,
a new approach was suggested by Khvedelidze, Lavrelashvili and Tanaka (KLT) \cite{Khvedelidze:2000cp}.
Using the full machinery of constrained dynamical systems, the quadratic action for the single physical degree of freedom was obtained, and in this framework the flat space limit is successfully recovered.
Note that similar results were later re-derived in \cite{Gratton:2000fj} and \cite{Dunne:2006bt}.
The end result of the KLT approach is similar to the earlier Lagrangian approach action,
but it contains  a different factor $Q$ in front of the kinetic term of the fluctuations, see Eq.~(\ref{eq:Q}) below.
It has been {\it proven} \cite{Khvedelidze:2000cp} that for bounces where $Q$ is everywhere positive there exists exactly one negative mode
in the spectrum of linear perturbations (see also \cite{Lavrelashvili:1999sr,Gratton:2000fj,Dunne:2006bt}).
In this case, in complete analogy with the non-gravitational case, one can interpret the instanton as mediating the decay of a metastable vacuum.

A recent study of gravitational instantons in flat potential barriers has highlighted the fact that for large classes of instantons, $Q$ becomes negative somewhere along the instanton \cite{Battarra:2013rba}. This has motivated us to study the issue of negative modes in some detail for these cases. Our findings are twofold:

First, we will demonstrate (analytically) that the eigenvalue equation retains non-singular solutions when $Q$ passes through zero. In other words, at $Q=0$ the equation is seemingly singular, but its solutions remain continuous and at least twice differentiable. Moreover, the eigenvalue evolves continuously as the potential is varied from cases where $Q>0$ everywhere to cases where $Q<0$ somewhere. In all the cases that we have investigated (numerically), the eigenvalue of the tunnelling negative mode remains negative. This supports the view that for CdL bounces the tunnelling negative mode always exists.

Second, we will confirm the existence of an infinite tower of additional negative modes when $Q$ becomes negative. They were already conjectured to exist by LRT in their early work \cite{Lavrelashvili:1985vn}, and here we provide some explicit numerical examples. These additional negative modes, which are always present in the region where $Q<0,$ remain puzzling: on the one hand, it does not seem possible to remove them by a canonical transformation (see appendix \ref{appCT}), but on the other hand the associated instantons do not appear to be physically significantly different from those where $Q>0$ everywhere. The existence of this additional set of modes remains somewhat mysterious and will require further clarification.

\section{Description of false vacuum decay in Euclidean approach}

Let's consider the theory of a single self-interacting scalar field
which is defined by the following Euclidean action
\be
S_E=\int {d^4x[\frac{1}{2} \partial_\mu\varphi\partial^\mu\varphi+V(\varphi)]} \pkt
\ee
Further, let's assume that $V(\varphi)$ is an asymmetric double-well potential and that it
has a local minimum (false vacuum) at some $\varphi=\varphi_f$, an absolute minimum
(true vacuum) at $\varphi=\varphi_{t}$ and a local maximum (top) at some $\varphi=\varphi_{top}$,
such that $\varphi_t < \varphi_{top} < \varphi_{f}$.

The energy $E_0$ of the lowest energy state localised around the false vacuum
gets a correction due to quantum tunnelling effects,
\be
E_f=E_0-\gamma \pkt
\ee
It turns out that the correction is purely imaginary $\gamma=i |\gamma|,$
which is a sign of metastability and shows that $\Gamma \equiv |\gamma|$ actually
describes the decay width of the false vacuum. This decay width is given by the functional integral
\be \label{fi1}
\gamma=\frac{1}{N_-}\int D\varphi \e^{-S_E(\varphi)}
\ee
where $N_-$ is a normalisation factor. In the quasi-classical approximation the functional integral Eq.~(\ref{fi1})
can be evaluated by considering small perturbations about
the classical saddle point Euclidean solution known as the ``bounce''.
The Euclidean action can be expanded as
\be
S_E=S^{(cl)}(\varphi^b)+S^{(2)}(\delta\varphi) |_{\varphi=\varphi^b} \pkt
\ee
The normalization factor is the same functional integral calculated about the
false vacuum $\varphi=\varphi_f$.
So, for $\gamma$ we find the Arrhenius formula
\be
\gamma = {\cal A} \e^{-{\cal B}} \kma
\ee
with
\be
B=S^{(cl)}(\varphi^b)-S^{(cl)}(\varphi^f)
\ee
and ${\cal A}$ being ratio of the corresponding integrals:
\be \label{eq:A}
{\cal A}= \frac
{\int D\delta\varphi \e^{ -S^{(2)}(\delta\varphi)}|_{\varphi=\varphi^b}}
{\int D\delta\varphi \e^{ -S^{(2)}(\delta\varphi)}|_{\varphi=\varphi^f}} \pkt
\ee
The quadratic action for the $O(4)$ symmetric configurations takes the form
\be \label{eq:qa_flat}
S_E^{(2)}=2\pi^2 \int \eta^3 d\eta \delta \varphi \hat{O}_F \delta \varphi \kma \quad {\rm with}~~~
\hat{O}_F = - \frac{1}{2\eta^3}\frac{d}{d\eta} \eta^3 \frac{d}{d\eta} +\frac{1}{2} V''(\varphi)\ .
\ee
So, the mode equation diagonalising the quadratic action Eq.~(\ref{eq:qa_flat}) has the form of a Schr\"odinger equation
\be \label{eq:flatSchroedinger}
[-\frac{d^2}{d\eta^2}-\frac{3}{\eta}\frac{d}{d\eta} + V''(\varphi)] \delta\varphi_n = \lambda_n \delta \varphi_n\ .
\ee
Since any perturbation with proper boundary conditions can be decomposed into a complete set of functions of the fluctuation operator $\hat{O}_F,$
\be
\delta\varphi = \sum_n c_n \delta \varphi_n \kma
\ee
integration over $\delta\varphi$ in Eq.~(\ref{eq:A}) can be replaced by integration over $c_n$.
Taking Gaussian integrals, one obtains the product of eigenvalues, i.e. determinants of the corresponding operators
\be
{\cal A} = \frac{B^2}{4\pi^2} \left(\frac{\displaystyle \prod_{n^\prime} \lambda_n^b}{\displaystyle \prod_n \lambda_n^f }\right)^{-\frac{1}{2}}
=\frac{B^2}{4\pi^2}
\left(\frac{det'[-\partial^2+V''(\varphi^b)]} {det[-\partial^2+V''(\varphi^{f})]}\right)^{-\frac{1}{2}} \pkt
\ee
Here a prime indicates that the zero-modes must be omitted - as described by Callan and Coleman \cite{Callan:1977pt}, a proper treatment of the zero-modes results in the $B^2/4\pi^2$ pre-factors. While finding bounce solutions and calculating the exponential factor in the thin-wall approximation (or numerically) is a relatively easy task,
the calculation of the pre-exponential factor in field theory is considerably more involved \cite{Baacke:2003uw,Dunne:2005rt,Dunne:2006bt} -- in particular, one has to deal with possible one-loop divergences.

\section{Bounce solutions with gravity}

Let's consider the theory of a single scalar field minimally coupled to gravity,
which is defined by the following Euclidean action
\be
S_E=\int {d^4x\sqrt{g} \; \Bigl(-\frac{1}{2\kappa}R + \frac{1}{2}\nabla_\mu\varphi \nabla^\mu\varphi+ V(\varphi) \Bigr)} \kma
\ee
where $\kappa=8\pi G_{N}$ is the reduced Newton's gravitational constant. The most general $O(4)$ invariant metric is parametrised as
\be\label{metric}
ds^2=N^2(\eta)d\eta^2+\rho^2(\eta)d\Omega_3^2 \kma
\ee
where $N(\eta)$ is the Lapse function, $\rho(\eta)$ is the scale factor
and $d\Omega_3^2$ is metric of the unit three-sphere,
\be
d\Omega_3^2 = d\chi^2 + {\rm sin}^2\chi (d\theta^2 + {\rm sin}^2(\theta) d\phi^2) \pkt
\ee
For the metric in Eq.~(\ref{metric}) the curvature scalar looks like
\be
R= \frac{6}{\rho^2}-\frac{6 \dot{\rho}^2}{\rho^2 N^2}-\frac{6 \ddot{\rho}}{\rho N^2}
+ \frac{6 \dot{\rho} \dot{N}}{\rho N^3} \kma
\ee
where $\dot{} = d/ d\eta$.
Using the ansatz (\ref{metric}) and assuming that $\varphi=\varphi(\eta),$
we get the reduced action in the form
\be
S_E=S_E(\varphi,N,\rho)= 2\pi^2 \int d\eta \Bigl(
\frac{\rho^3}{2N}\dot{\varphi}^2 + \rho^3 N V(\varphi)
-\frac{3 \rho N}{\kappa} +\frac{3 \rho \dot{\rho}^2}{\kappa N}
+\frac{3 \rho^2 \ddot{\rho}}{\kappa N}
-\frac{3 \rho^2 \dot{\rho}\dot{N}}{\kappa N^2} \Bigr) \pkt
\ee
In proper-time gauge, $N=1,$ the corresponding field equations are
\be \label{eq:phi}
\ddot{\varphi}+3\frac{\dot{\rho}}{\rho}\dot{\varphi}=\frac{\partial V}{\partial\varphi} \kma
\ee
\be \label{eq:rho}
\ddot{\rho}=-\frac{\kappa \rho}{3} (\dot{\varphi}^2+V(\varphi)) \kma
\ee
\be
\dot{\rho}^2= 1+ \frac{\kappa \rho^2}{3}(\frac{\dot{\varphi}^2}{2}-V) \pkt
\ee
Now let's assume that the potential $V(\varphi)$ has two non-degenerate local minima
at $\varphi=\varphi_{\rm t}$ and $\varphi=\varphi_{\rm f}$, with $V(\varphi_{\rm f})>V(\varphi_{\rm t})$,
and a local maximum for some $\varphi=\varphi_{\rm top}$, with $\varphi_{\rm t}<\varphi_{\rm top}<\varphi_{\rm f}$.
The Euclidean solution describing vacuum decay - the bounce - satisfies
these equations, and when $V(\varphi) > 0$ one has the boundary conditions
\be \label{eq:initial_conditions}
\varphi (0)= \varphi_0,\qquad \dot{\varphi}(0) = 0,\qquad \rho(0)=0, \qquad \dot{\rho}(0)=1
\ee
at $\eta=0$ and
\be \label{eq:etamax_conditions}
\varphi (\eta_{max})= \varphi_{m},\qquad \dot{\varphi}(\eta_{max}) = 0,\qquad
\rho(\eta_{max})=0,\qquad \dot{\rho}(\eta_{max})=1
\ee
at some $\eta=\eta_{max}$.
This assumes the following Taylor series as $\eta \to 0$:
\bea \label{eq:taylor_phi}
\varphi(\eta)&=&\varphi_0 +\frac{V'(\varphi_0)}{8} \eta^2
+\frac{V'(\varphi_0)}{192} [ V''(\varphi_0) +\frac{2\kappa V(\varphi_0)}{3}] \eta^4
+ \frac{V'(\varphi_0)}{829440}  \bigl[ 135 V'(\varphi_0) V'''(\varphi_0)  \bigr. \nn \\
\bigl.
&+&90 V''(\varphi_0)^2 + 162 \kappa V'(\varphi_0)^2 +180 \kappa V(\varphi_0) V''(\varphi_0) +112 \kappa^2 V(\varphi_0)^2 \bigr] \eta^6  + O(\eta^8) \kma \\
\label{eq:taylor_rho}
\rho(\eta)&=&\eta-\frac{\kappa}{18}V(\varphi_0)\eta^3
-\frac{\kappa}{120}[\frac{3}{8} V'(\varphi_0)^2 -\frac{\kappa}{9}V(\varphi_0)^2] \eta^5 \nn \\
&-&\frac{\kappa}{2177280} \bigl[ 405 V'(\varphi_0)^2 V''(\varphi_0)-54 \kappa V(\varphi_0) V'(\varphi_0)^2 +16 \kappa^2 V(\varphi_0)^3 \bigr] \eta^7
+ O(\eta^9) \kma
\eea
where $V'(\varphi_0)\equiv \frac{\partial V}{\partial \varphi}|_{\varphi=\varphi_0}$ etc.
Similar power-law behavior is valid for non-singular bounces for $x\to 0$, where $x=\eta_{max}-\eta$. These Taylor series are required when numerically solving for instanton solutions, as one cannot directly integrate from $\eta = 0$ (given the above boundary conditions), but rather has to start the integration at some small value $\eta = \epsilon \ll 1.$

\section{Negative mode problem in Hamiltonian approach} \label{sec:neg_mode}

Let's expand the metric and the scalar field over a $O(4)-$symmetric background as follows:
\be \label{eq:homogen_metric}
ds^2=  (1+2 A(\eta))d\eta^2 + \rho(\eta)^2 (1-2 \Psi(\eta))d\Omega_3^2
\kma \qquad \varphi=\varphi(\eta) + \Phi(\eta) \kma
\ee
where $a$ and $\varphi$ are the background field values and
$A, \Psi$ and $\Phi$ are small perturbations.
In what follows, we will be interested in the lowest (purely $\eta$-dependent, `homogeneous') modes
and consider only scalar metric perturbations. Expanding the total action to second order in perturbations
and using the background equations of motion, we find
\be
S= S^{(0)}[a,\varphi]+S^{(2)}[A,\Psi,\Phi] \kma
\ee
where $S^{(0)}$ is the action of the background solution and $S^{(2)}[A,\Psi,\Phi]$
is the quadratic action given below.

The quadratic action about CdL bounces was first derived in \cite{Lavrelashvili:1985vn} using the Lagrangian approach
and in particular it was noted that when gravity is taken into account the corresponding operator $\hat{O}_G$
in front of the kinetic term contains a factor (see the Appendix for more details)
\be \label{eq:Qlrt}
Q_{LRT} = 1-\frac{\kappa \rho^2 V(\varphi)}{3}=\dot{\rho}^2 -\frac{\kappa \rho^2 \dot{\varphi}^2}{6} \pkt
\ee
This factor typically becomes negative somewhere along a bounce solution. Later, using a Hamiltonian approach in the context of the theory of constrained dynamical systems,
the quadratic action was rederived \cite{Khvedelidze:2000cp} and it was shown to have a similar structure, but
with a different factor in front of the kinetic term of the perturbations,
\be \label{eq:Q}
Q_{KLT}\equiv Q = 1-\frac{\kappa \rho^2 \dot{\varphi}^2}{6} \pkt
\ee

The original quadratic action $S^{(2)}[A,\Psi,\Phi] $ is degenerate and describes a constrained dynamical system.
Applying Dirac's formalism as in \cite{Khvedelidze:2000cp}, one gets an unconstrained quadratic action for a single physical degree
of freedom (which for this gauge fixing procedure is simply the scalar field perturbation $\Phi$)
\be \label{qa}
S^{(2)}_E[\Phi] = \pi^2 \int d\eta \Phi \Bigl(-\frac{d}{d\eta} (\frac{\rho^3(\eta)}{Q(\eta)}\frac{d}{d\eta})
+ \rho^3(\eta) U[\varphi(\eta),\rho(\eta)] \Bigr) \Phi \kma
\ee
where the factor $Q$ was given above in Eq.~(\ref{eq:Q}) and the potential $U$ is expressed in terms of the bounce solution as
\be \label{eq:Hamiltonian_U}
U[\varphi(\eta),\rho(\eta)] \equiv \frac{V''(\varphi)}{Q}+\frac{2\kappa{\dot{\varphi}}^2}{Q}
+\frac{\kappa}{3 Q^2} \Bigl(6{\dot{\rho}}^2{\dot{\varphi}}^2 +\rho^2 V'^2(\varphi)
-5\rho \dot{\rho}\dot{\varphi} V'(\varphi)\Bigr) \pkt
\ee
The exact form of the fluctuation operator depends on the choice of a weight function,
which can be specified by defining a norm. In the context of general relativity the natural choice is \cite{Dunne:2006bt}
\be
|| \Phi ||^2 \equiv \int d^4 x \  \sqrt{g} \ \Phi^2 = 2 \pi^2 \int d\eta \ \rho(\eta)^3 \ \Phi^2 \;.
\ee
The fluctuation equation diagonalizing the quadratic action Eq.~(\ref{qa}) then has the form
\be \label{eq:fe}
-\frac{1}{\rho^3}\frac{d}{d\eta}(\frac{\rho^3}{Q}\frac{d\Phi_n}{d\eta}) + U[\varphi(\eta),\rho(\eta)] \Phi_n = \lambda_n \Phi_n \kma
\ee
where $\Phi_n$ and $\lambda_n$ are eigenfunctions and eigenvalues of our Dirichlet boundary value problem. Note that this equation correctly reduces to the flat-space equation \eqref{eq:flatSchroedinger} in the limit $\kappa \rightarrow 0$, $\rho \rightarrow \eta.$

The potential $U$, close to $\eta=0,$ behaves as
\be
U= U_0 + U_2 \ \eta^2 + O[\eta^4] \kma
\ee
with
\be
U_0= V''(\varphi_0) \kma~~~{\rm and}~~~
U_2=\frac{1}{6}\kappa V'^2(\varphi_0) + \frac{1}{8}V'(\varphi_0) V'''(\varphi_0) \pkt
\ee
Regular solutions (eigenfunctions) close to $\eta=0$ then behave as
\bea
\Phi &=& A_0 \{ 1+\frac{1}{8}(V''(\varphi_0)-\lambda) \ \eta^2 +
\frac{1}{576}[ 3 (V''(\varphi_0)-\lambda)^2+ 2\kappa (V''(\varphi_0)-\lambda) V(\varphi_0)  \nn \\
 &+&4 \kappa V'^2(\varphi_0) +3 V'(\varphi_0) V'''(\varphi_0) ] \ \eta^4 + O[\eta^6] \} \kma
\eea
with $A_0$ being a normalisation constant. Obviously, the potential $U$ and the regular branch of the wavefunctions $\Phi^{reg}$ have the same behaviour in powers of $x = \eta_{max}-\eta$
at the end of the interval, close to $\eta_{max}$.
Whereas in general there exists also a {\it singular} branch behaving as $\Phi^{sing} \propto \frac{f(\lambda-\lambda_n)}{x^2}$ with some function $f$
having the property that $f(0)=0$, one can adjust the value of $\lambda$ such that the singular branch is suppressed. In fact, in this way one may determine the (quantised) energy eigenvalues of the eigenfunctions.

\section{Regularity of the perturbations when $Q$ passes through a zero} \label{section:Regularity}

As was discussed in Sect.~\ref{sec:neg_mode}, the equation for linear perturbations
in the Hamiltonian approach has the form
\be \label{eq:perturbD}
- \frac{1}{Q} \frac{d^2 \Phi}{ d\eta^2}
+\left(\frac{\dot{Q}}{Q^2} - \frac{3 \dot{\rho}}{ \rho Q}\right) \frac{d \Phi}{d\eta} + U  \Phi = \lambda \Phi \kma \\
\ee
with the potential $U$ given in Eq.~(\ref{eq:Hamiltonian_U}). Note that the function $Q$ tends to 1 at the ends of the interval $[0, \eta_{max}]$,
but for some bounces it can become negative for some interval of $\eta$. The case where $Q$ becomes negative has long been considered as puzzling \cite{Khvedelidze:2000cp,Gratton:2000fj,Dunne:2006bt} because Eq.~(\ref{eq:perturbD}) then shows an {\it apparent} singularity when $Q$ passes through zero.
We will now show that when $ Q( \bar{ \eta}) = 0$  the general solution of \eqref{eq:perturbD} is nevertheless $ \mathcal{C} ^{2}$
across the point $ \bar{ \eta}$. We treat the cases where $\dot{Q}(\bar{ \eta}) \neq 0$ and $ \dot{Q}(\bar{ \eta}) = 0$ separately.
We will assume that $Q$ and the other background quantities are at least $ \mathcal{C} ^2$ functions of $ \eta$.

\subsection{$\dot{Q}(\bar{\eta})\neq 0$ case}

Let $x \equiv \eta - \bar{ \eta}$. We can re--write \eqref{eq:perturbD} as
\be \label{eq:fullApproxD}
- Q\, \ddot{\Phi} + \left( \dot{Q} - 3 \frac{ \dot{\rho}}{ \rho} Q \right) \dot{\Phi} + \mathcal{U} \Phi =\lambda Q^2 \Phi \kma \\
\ee
with
\be \label{eq:U}
\mathcal{U}  \equiv  Q^2 U = Q\, V'' + 2 \kappa Q \dot{\varphi}^2 + \frac{ \kappa \rho ^2 V'^2}{3}
- \frac{ 5 \kappa \rho \dot{\rho} \dot{\varphi} V'}{3} + 2 \kappa \dot{\rho}^2 \dot{\varphi}^2  \pkt
\ee
We start with the consideration of the zero mode equation, i.e. we assume $\lambda=0$ in  Eq.~(\ref{eq:fullApproxD}).
Since we are interested in showing the regularity of the wave function
in the vicinity of $x=0$, we re--write \eqref{eq:fullApproxD} as
\begin{equation}
 - x A(x) \, \ddot{\Phi} + B(x)\, \dot{\Phi} + \frac{ u}{ \dot{q}} C(x) \Phi = 0 \;,
\end{equation}
where we denote
\begin{eqnarray}
\dot{q} & \equiv & \dot{Q}( \bar{ \eta}), \quad \ddot{q} \equiv  \ddot{Q}( \bar{ \eta}), \ldots\;, \\
u & \equiv & \mathcal{U}( \bar{ \eta}), \quad \dot{u} \equiv \dot{\mathcal{U}}( \bar{ \eta}), \ldots \;,\\
A(x) & = & 1 + \mathcal{O}(x) \;,\\
B(x) & = & 1 + b\,x + \mathcal{O}(x ^2), \quad b = \frac{\ddot{q}}{ \dot{q}} - 3 \frac{ \bar{ \dot{\rho}}}{ \bar{ \rho}} \;, \\
C(x) & = & 1 + c\,x + \mathcal{O}(x ^2), \quad c = \frac{\dot{u}}{ u} \;.
\end{eqnarray}
We tentatively solve the equation by keeping the dominant behaviour of the coefficients first,
\begin{equation} \label{eq:firstApproxD}
- x\, \ddot{\Phi} + \dot{\Phi} +  \frac{ u}{ \dot{q}} \Phi = 0 \;.
\end{equation}
The general solution of this equation can be written in terms of Bessel functions. The leading behaviour of the solution is of two different kinds:
\begin{equation} \label{eq:solFirstApproxD}
\Phi = \alpha  \left( 1 - \frac{u}{ \dot{q}} x - \frac{u ^2}{ 2 \dot{q}^2 } x ^2 \ln{|x|} + \ldots \right) + \beta\, x ^2  (1 + \ldots ) \;.
\end{equation}
However, the second sub--leading term in the first asymptotic solution is not relevant, since it contributes a term of order $ x$ to the perturbation equation, which is the same as the dominant order we neglected in going from \eqref{eq:fullApproxD} to \eqref{eq:firstApproxD}. To find the right expression for the sub--leading term and make sure that $ \ddot{\Phi}$ is well defined at $ x=0$, we set
\begin{equation}
\Phi  = \alpha \left( 1 - \frac{u}{\dot{q}}x + x\, \psi \right) \;.
\end{equation}
The asymptotic equation for $ \psi$ reads
\begin{equation} \label{eq:approxOnceMore}
- x ^2 \ddot{\psi} - x \dot{\psi} + \psi + \frac{u}{\dot{q}}\left( c - b - \frac{u}{ \dot{q}} \right)x = 0\;.
\end{equation}
Discarding the solution $ \sim 1/x$, which simply changes the value of $ \alpha$, the leading behaviour of the solution is the same as in \eqref{eq:solFirstApproxD}, albeit with a different coefficient:
\begin{equation} \label{eq:finalExpressionNonDegenerate}
\Phi = \alpha \left(1 - \frac{u}{\dot{q}} x + \frac{u}{2 \dot{q}} \left( c - b - \frac{ u}{\dot{q}} \right) x ^2 \ln{|x|}
+ \ldots \right) + \beta\, x ^2 ( 1+ \ldots ) \;.
\end{equation}
However, a closer analysis of the non--homogeneous piece appearing in \eqref{eq:approxOnceMore} shows that its coefficient vanishes due to the background field equations. This means that one should move to the next--to leading order for the non--homogeneous piece, and the solution for $ \psi$ is of order $ x$ with no logarithmic corrections:
\begin{equation}
\Phi = \alpha \left( 1 - \frac{u}{\dot{q}}x + \mathcal{O}(x ^2) \right) + \beta x ^2 ( 1 + \ldots) \;.
\end{equation}
This completes our proof: the apparent singularity in the perturbation equation forces a fixed relation between $ \Phi( \bar{ \eta})$ and
$ \dot{\Phi}( \bar{ \eta})$, but does not cause any divergence in the solution, at least for $ \Phi$ and its first two derivatives.

We now note that, if one considers the eigenvalue equation (\ref{eq:perturbD}) with $\lambda \neq 0$
the result above is still valid, since this merely amounts to a change in $ C(x)$ by a term of order $ x ^2$. Therefore, the generic solution of the eigenvalue equation is also non--singular at $ \eta = \bar{ \eta}$.

\subsection{$ \dot{Q} ( \bar{ \eta}) = 0$ case}

In this case, we can't simply borrow the previous results, as $\dot{q} = 0$. 
The field equations alone do not determine the leading degree of the $ Q$ polynomial around $ \eta =  \bar{ \eta}$. 
Keeping it general, we will write
\begin{equation} \label{eq:asyQ}
Q = \frac{q ^{(k)}}{ k!}x ^{k} + \frac{ q ^{(k+1)}}{ (k+1)!} x ^{k+1} + \mathcal{O}(x ^{k+2}), \quad k>1 \;.
\end{equation}
We require $k>1$ to be an integer.
At first look, one would think that $ \mathcal{U} \equiv ( Q ^2 U)$ should not generally vanish at $ \eta = \bar{ \eta}$, so that the leading terms of the perturbation equation would read
\begin{equation}
- x ^{k} \ddot{\Phi} + k\, x ^{k-1} \dot{\Phi} + k!\, \frac{u}{ q ^{(k)}}\, \Phi  = 0 \;,
\end{equation}
where again $u \equiv \mathcal{U}( \bar{ \eta})$ is generally non--zero. This would be a problem, as the equation above admits solutions which diverge near $ \eta = \bar{ \eta}$. A closer analysis shows that, in fact, the behaviour of $\mathcal{U}$ near $ \bar{ \eta}$ is related to that of $ Q$ itself.
Let us show that, under the condition \eqref{eq:asyQ}, one has
\begin{equation} \label{eq:requiredBound}
| \mathcal{U}| \lesssim |x| ^{k-1} \;.
\end{equation}
We have
\begin{eqnarray} \label{eq:qsq}
\mathcal{U} & = & Q \left( V'' + 2 \kappa \dot{\varphi}^2 \right) + \bar{u}  \;,\\
\bar{u}  & \equiv & \frac{ \kappa \rho ^2 V'^2}{3}  - \frac{ 5 \kappa \rho \dot{\rho} \dot{\varphi} V'}{3} + 2 \kappa \dot{\rho}^2 \dot{\varphi}^2 \;.
\end{eqnarray}
The first two terms in \eqref{eq:qsq} separately satisfy the bound \eqref{eq:requiredBound}. Concerning the last piece, using the background field equations one can show that
\begin{equation} \label{eq:baru}
\bar{u} = \frac{ \dot{\rho} \dot{Q}}{ \rho} + \frac{ 3 \dot{Q} ^{2}}{ \kappa \rho ^2 \dot\varphi ^{2}} \;.
\end{equation}
This relation is valid everywhere, not only at $ \eta = \bar{ \eta}$. Taking its derivatives up to order $k-2$,
one easily deduces that (remembering that $\dot{\varphi}(\bar{\eta})\neq 0$ since $Q(\bar{\eta})=0$)
\begin{equation}
\bar{u} ^{(n)}( \bar{ \eta}) = 0: \quad n = 0, \ldots, k-2 \;,
\end{equation}
and hence \eqref{eq:requiredBound},
\begin{equation}
\mathcal{U} = \frac{\bar{u} ^{(k-1)}}{(k-1)!} x^{k-1} + \frac{\bar{u} ^{(k)}}{k!} x ^{k} + \ldots \;.
\end{equation}
Now, dividing through by $x^{k-1}$ and keeping the leading terms near $\eta = \bar{\eta},$ we obtain an equation very similar to Eq. (\ref{eq:firstApproxD}), namely
\begin{equation} \label{eq:firstApproxDQdotzero}
- x\, \ddot{\Phi} + k \dot{\Phi} + k \frac{ \bar{u}^{(k-1)}}{ q^{(k)}} \Phi = 0 \;.
\end{equation}
 The asymptotic form of the general solution can be obtained following the previous procedure, finding:
\begin{eqnarray}
\Phi & = & \alpha \left( 1- \frac{\bar{u} ^{(k-1)}}{q ^{(k)}}x + \frac{d}{2}\, x ^2 + \ldots \right) + \beta x ^{k+1} \left(1 + \ldots \right) \;,\\
d & = & \frac{\bar{u} ^{(k-1)}}{(k-1) q ^{(k)}} \left( k \frac{ \bar{u} ^{(k-1)}}{q ^{(k)}} + \frac{q ^{(k+1)}}{q ^{(k)}} - 3 \frac{ \dot{\bar{ \rho}}}{ \bar{ \rho}} \right) - \frac{ \bar{u} ^{(k)}}{(k-1) q ^{(k)}}  \;.
\end{eqnarray}
Note that in this case (i.e. for $k>1$), the $x^2 \ln |x|$ term right away does not appear in the expansion.\footnote{We also note that the term linear in $x$ may be absent if $\bar{u}^{(k-1)}=0$. Eq. \eqref{eq:baru} shows that this can happen if $\frac{ \dot{\bar{ \rho}}}{ \bar{ \rho}}=0,$ which in turn can only happen when $V^\prime(\bar\eta)=0,$ as can be seen by combining the expression for $\dot{Q}$ and the scalar field equation of motion.} We conclude that also in this case, the apparent singularity in the perturbation equation does not lead to a singularity in the general solution.

\section{Numerical results}

We have studied the existence and the properties of fluctuation modes about CdL bounces numerically for a range of representative potentials. We have mostly used the same potentials as those used by Lee and Weinberg in their related study \cite{Lee:2014uza}, as this allows us to better compare our results with theirs (see appendix \ref{appLW}).

In order to find a bounce solution we have to integrate Eqs.~(\ref{eq:phi}, \ref{eq:rho}) with the initial conditions
(\ref{eq:initial_conditions}). But since Eq.~(\ref{eq:phi}) has a regular singular point at $\eta=0$ we cannot start the numerical integration there.
Instead we start the numerical integration at some small $\eta=\epsilon$, with initial conditions at this point provided by the Taylor series
Eqs.~(\ref{eq:taylor_phi}), (\ref{eq:taylor_rho}). This is a one-dimensional shooting problem with the shooting parameter being the
initial value of the scalar field, $\varphi_0$. For the compact bounces under consideration, one has to choose $\varphi_0$ such that at $\eta=\eta_{max}$
one gets the behaviour given by Eqs.~(\ref{eq:etamax_conditions}).
Since the initial conditions are given with finite precision, at some $\eta<\eta_{max}$ any such numerical solutions will start to deviate from the exact solution,
because of a mixture with the singular branch. For some potentials, one needs to adjust the initial values to very high precision to construct the bounce solution.
The presented Taylor series of Eqs.~(\ref{eq:taylor_phi})--(\ref{eq:taylor_rho}) guarantee precision of about $10^{-40}$ for $\epsilon=10^{-5}$. In the following, we specify different potentials and report our numerical results regarding concrete examples for various potentials and varying values of $\kappa$. The numerical method that we employed is the Runge-Kutta of fourth order with an adaptive step size for the background quantities and with a fixed step size for the mode functions, and we implemented it in the C++ programming language.

\subsection{Example 1}

\begin{figure}
\subfigure[~~$\kappa=0.055$]{\includegraphics[width=.495\textwidth]{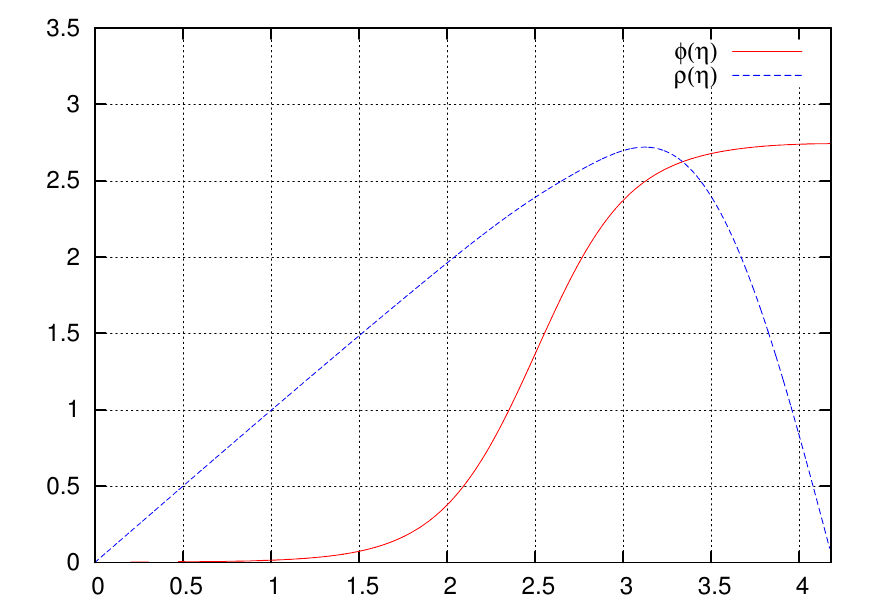}\label{fig:lw1-55}}
\subfigure[~~$\kappa=0.09$]{\includegraphics[width=.495\textwidth]{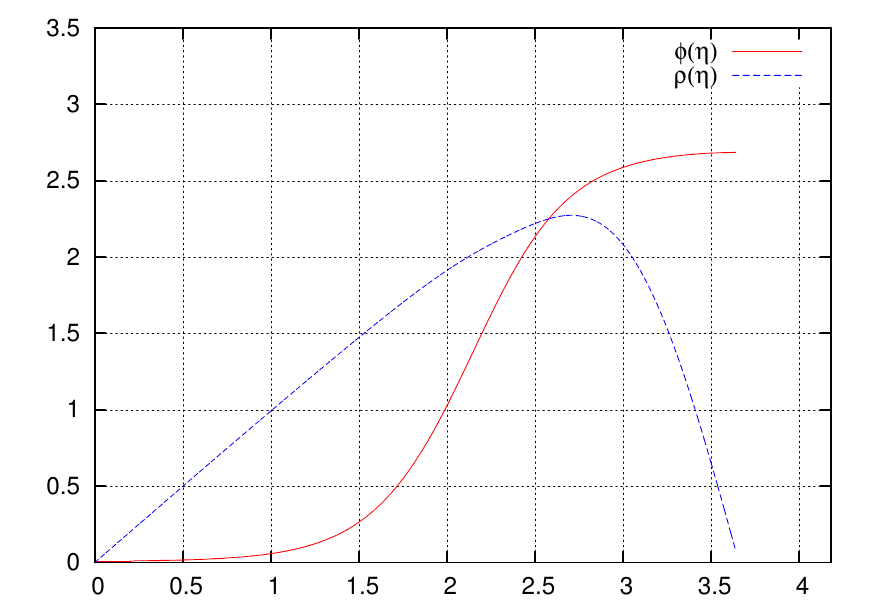}\label{fig:lw1-9}}
\caption{Profile of the bounce: $\varphi(\eta)$ and $\rho(\eta)$ for the potential (\ref{eq:LW1}) and two different values of $\kappa$.}
\label{fig:lw1}
\end{figure}
\begin{figure}
\subfigure[~~$\kappa=0.055$ and $\kappa=0.057$]{\includegraphics[width=.495\textwidth]{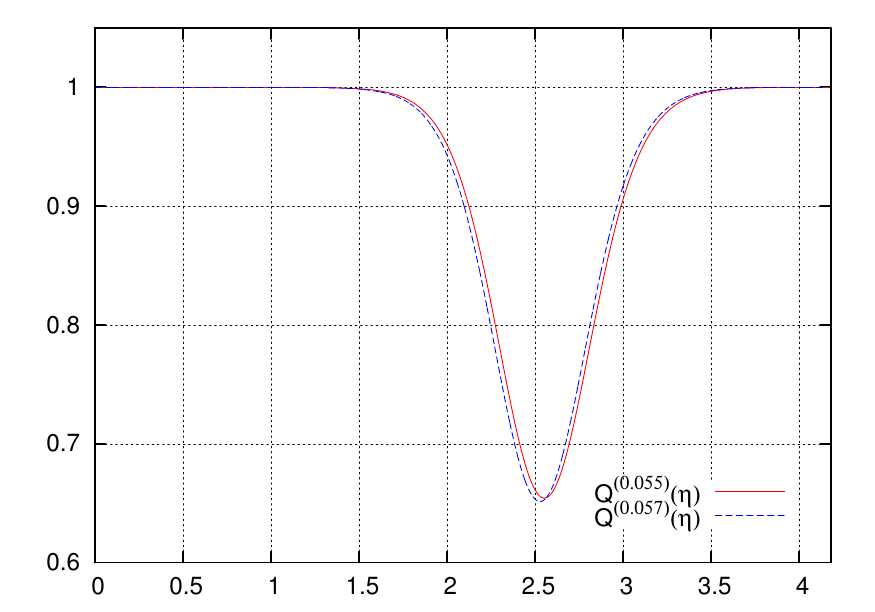}}
\subfigure[~~$\kappa=0.07$ and $\kappa=0.09$]{\includegraphics[width=.495\textwidth]{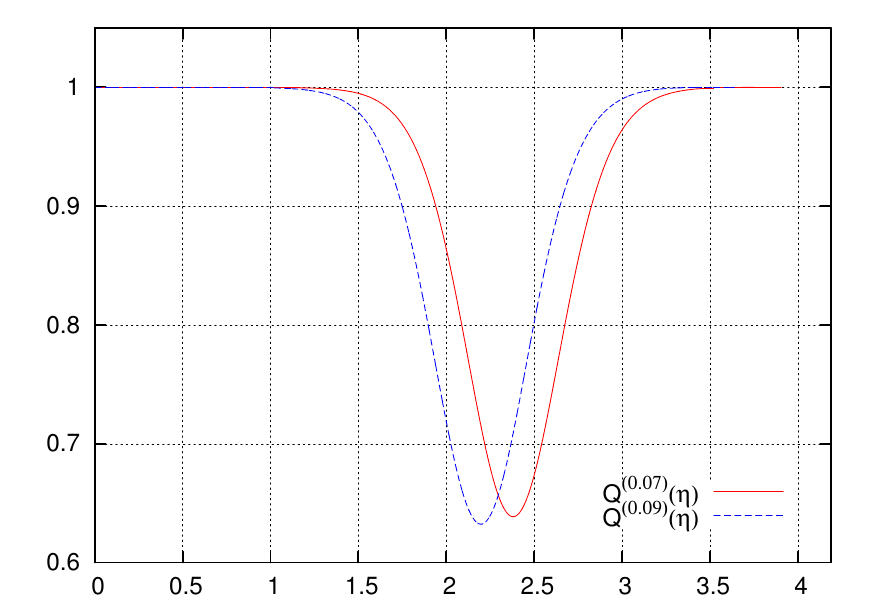}}
\caption{The kinetic pre-factor $Q(\eta)$ for the potential (\ref{eq:LW1}) and different values of $\kappa$.}
\label{fig:Q-lw1}
\end{figure}
\begin{figure}
\includegraphics[width=.495\textwidth]{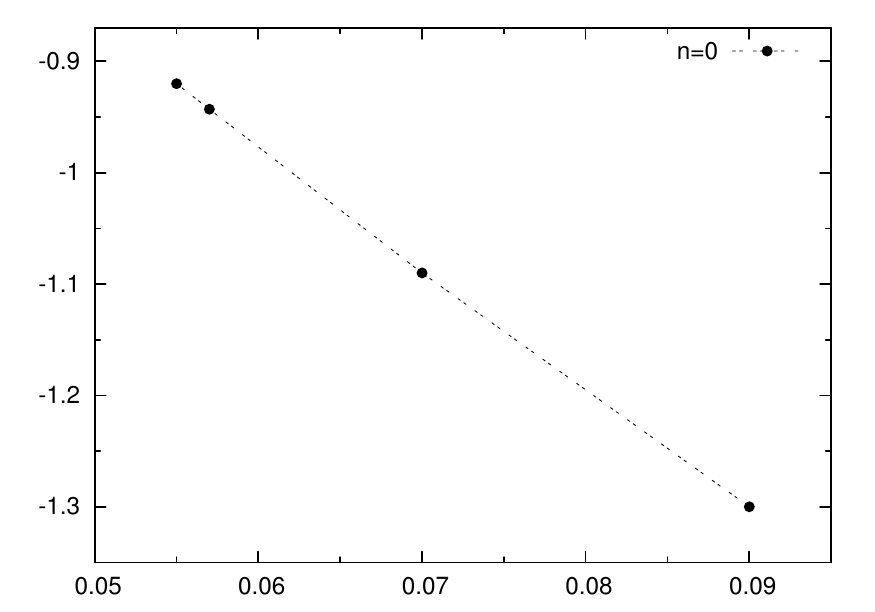}
\caption{Lowest eigenvalues $\lambda_0$ of the fluctuation equation (\ref{eq:fe}) versus $\kappa$ for the potential (\ref{eq:LW1}).}
\label{fig:lambda-kappa}
\end{figure}

As a first example, we have chosen the double-well potential also studied in \cite{Lee:2014uza}, namely
\be \label{eq:LW1}
V(\varphi) = (\varphi-3)^2 \varphi^2+\frac{1}{2} \varphi^2 + \frac{3}{2} \pkt
\ee
In this potential the true vacuum at $\varphi_t = 0 $ and the false vacuum at $\varphi_f \approx 2.81$
are separated by a potential barrier with local maximum at $\varphi_{top} \approx 1.69 $.

We are interested in bounce solutions in this potential, but for different values of the gravitational coupling $\kappa.$
Fig.~\ref{fig:lw1-55} shows the CdL bounce solution for this potential when $\kappa=0.055$,
while Fig.~\ref{fig:lw1-9} shows the bounce solution for $\kappa=0.09$.
It was claimed in \cite{Lee:2014uza} that a single tunnelling negative mode is present for small
values of $\kappa$ such as $\kappa = 0.055$, but that no negative mode is present for larger values,
namely for $\kappa = 0.057$, $\kappa=0.07$ or $\kappa=0.09$.
By contrast, we find that a single tunnelling negative mode exists in each of these cases.
Note that the quantity $Q$ is always positive for these choices of $\kappa$, see Fig.~\ref{fig:Q-lw1}, and
that the (negative) eigenvalue of the lowest mode evolves continuously as a function of the gravitational strength $\kappa,$ as shown in
Fig.~\ref{fig:lambda-kappa}. Thus, for this example, the story unfolds just as in the case of tunnelling in the absence of gravity.

\subsection{Example 2}

\begin{figure}
\subfigure[~$B=3$]{\includegraphics[width=.495\textwidth]{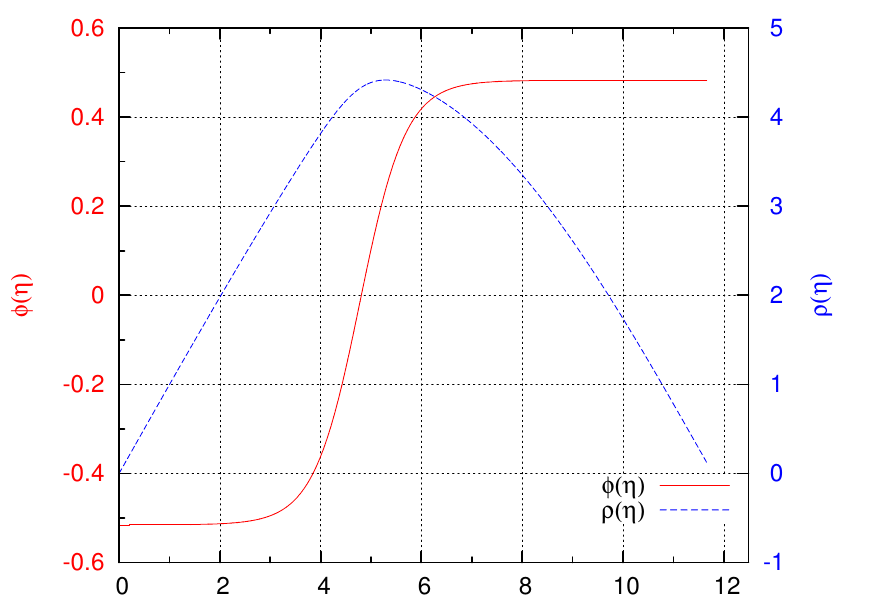}\label{fig:lw2-3}}
\subfigure[~$B=10$]{\includegraphics[width=.495\textwidth]{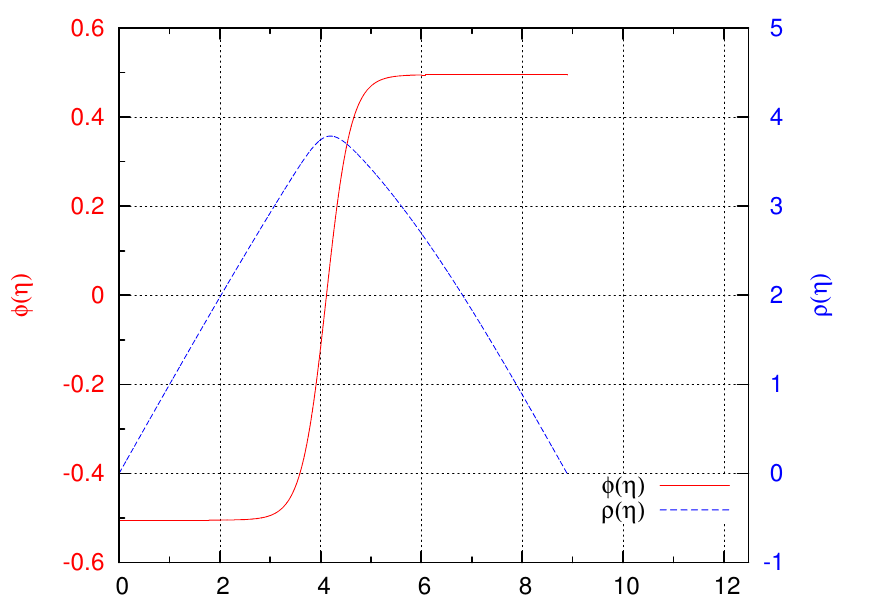}\label{fig:lw2-10}}
\caption{Profile of the bounce: $\varphi(\eta)$ and $\rho(\eta)$ for the potential (\ref{eq:LW2}) and different choices of $B$.}
\label{fig:lw2}
\end{figure}

\begin{figure}
\subfigure[~$B=2$ and $B=3$]{\includegraphics[width=.325\textwidth]{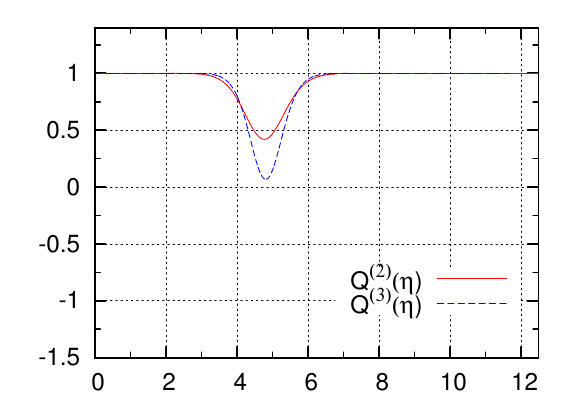}}
\subfigure[~$B=3.22$ ($Q<0$ briefly)]{\includegraphics[width=.325\textwidth]{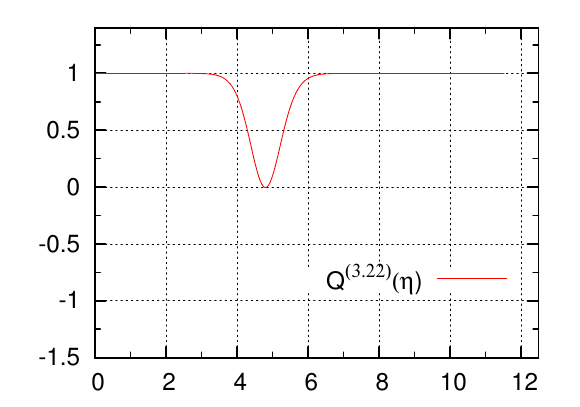}}
\subfigure[~$B=5$ and $B=10$]{\includegraphics[width=.325\textwidth]{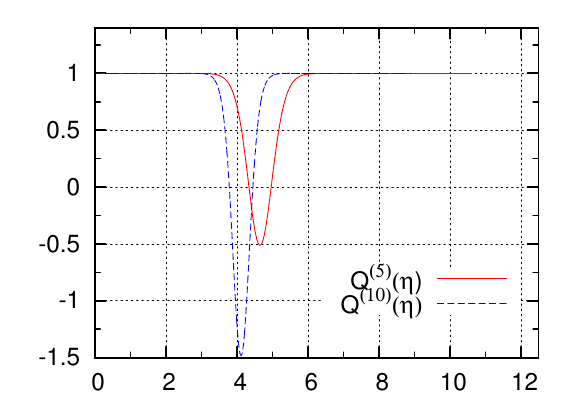}}
\caption{The kinetic pre-factor $Q(\eta)$ for the potential (\ref{eq:LW2}) and different values of $B$. For $B \gtrapprox 3.22$ $Q$ becomes negative along some interval of Euclidean time $\eta$.}
\label{fig:Q-lw2}
\end{figure}

\begin{figure}
\subfigure[~$\lambda_1$ and $\lambda_2$]{\includegraphics[width=.495\textwidth]{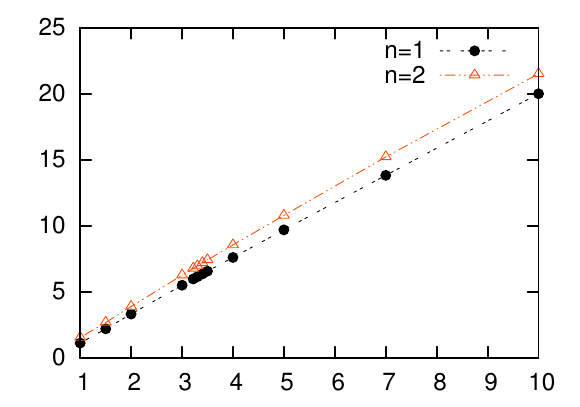}}
\subfigure[~$\lambda_0$]{\includegraphics[width=.495\textwidth]{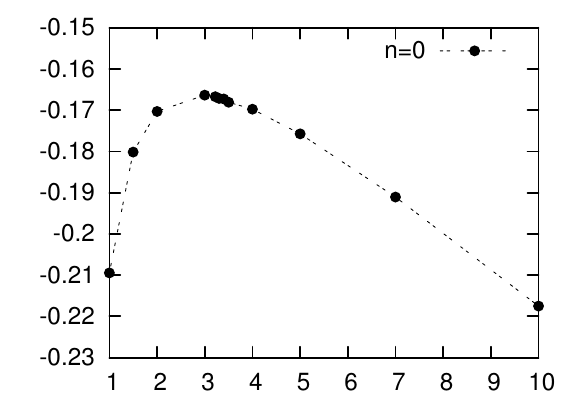}}
\caption{The first two positive eigenvalues $\lambda_1$ and $\lambda_2$ and the first (tunnelling) negative eigenvalue $\lambda_0$
of the fluctuation equation (\ref{eq:fe}) versus $B$ for the potential (\ref{eq:LW2}).}
\label{fig:table}
\end{figure}

\begin{figure}
\subfigure[~$\psi_1(\eta)$ and $\psi_2(\eta)$]{\includegraphics[width=.495\textwidth]{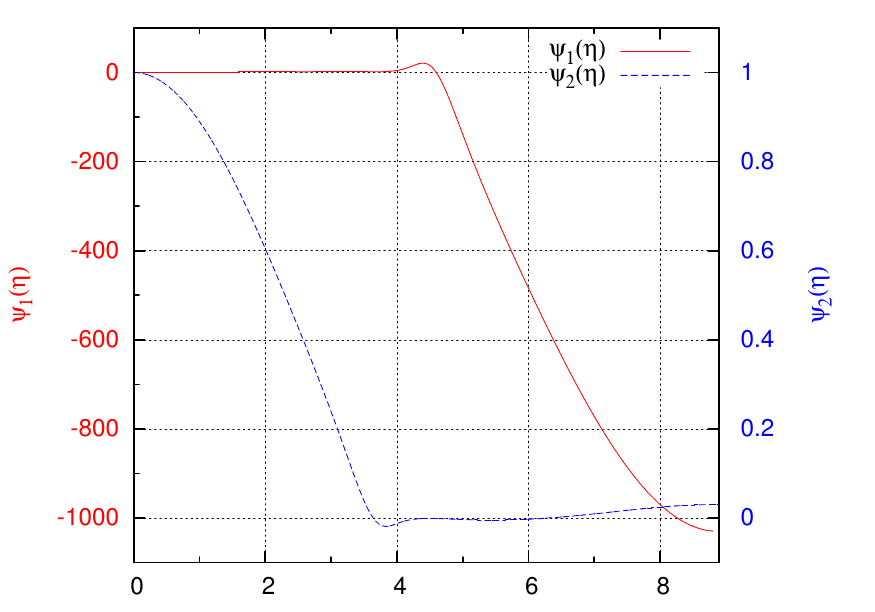}}
\subfigure[~$\psi_0(\eta)$]{\includegraphics[width=.495\textwidth]{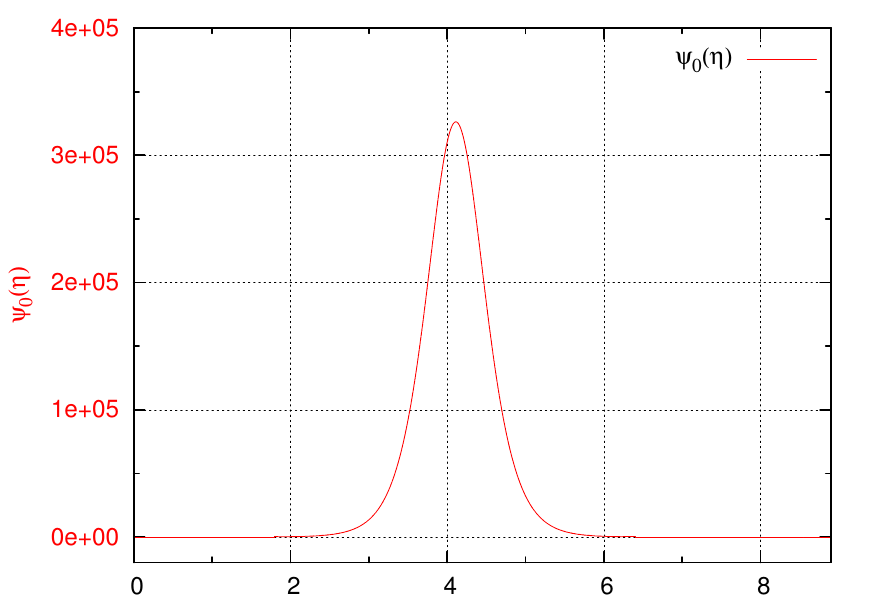}}\\
\caption{The wave functions of the first two excited positive modes and of the tunnelling negative mode
in the case of $B=10$ for the potential (\ref{eq:LW2}). Note the separate axes for the two positive modes on the left and right of graph (a).}
\label{fig:modesB10}
\end{figure}

As the second example we choose the potential (cf. \cite{Lee:2014uza})
\be \label{eq:LW2}
V(\varphi) = B (\varphi^2 - \frac{1}{4})^2  + \frac{1}{10} (\varphi + 1) \pkt
\ee
This potential has two local minima separated by a barrier, as long as the parameter $B>B_{cr}$, with $B_{cr} \approx 0.52$. A
CdL bounce exists for $B> B_{HM}$, where $B_{HM} \approx 0.55$.
Varying the parameter $B$ is instructive because for the potential \eqref{eq:LW2},
the factor $Q$ is always positive along the bounce trajectory in the case of small values of $B$,
while for $B>B_0$ with $B_0 \approx 3.22$, the bounce solution develops a region of negative $Q$. Therefore this potential is ideally suited for studying the case of principal interest here, namely what happens when $Q$ becomes negative.
We analysed the solutions for selected parameter values $B$ varying between $1$ and $10$.
Our main finding is that nothing special happens with the tunnelling negative mode when $Q$ becomes negative (as can now be expected, given the analytic treatment in section \ref{section:Regularity}).
Fig.~\ref{fig:lw2-3} shows the bounce profile for $B=3$, while
Fig.~\ref{fig:lw2-10} displays the corresponding bounce when $B=10$.
Note that factor $Q(\eta)$ is always positive for $B < B_0$ and is negative in some interval of $\eta$
for $B > B_0$, see Fig.~\ref{fig:Q-lw2}.
In this setup we investigated the tunneling negative mode and the first few positive modes in detail.
Fig.~\ref{fig:table} illustrates the dependence of the eigenvalue of the tunnelling negative mode $\psi_0$ as well
as the first two excited positive modes $\psi_{1,2}$ on the value of $B$.
The characteristic profiles of these modes are shown in Fig.~\ref{fig:modesB10} for the particular value of $B=10$. Note that the eigenvalues evolve continuously as a function of $B$ and in particular that the eigenvalue of the tunnelling negative mode always remains negative, regardless of whether $Q$ is positive everywhere or negative somewhere.


\begin{figure}
\subfigure[~$\psi_{-1}(\eta)$ and $\psi_{-2}(\eta)$]{\includegraphics[width=.6\textwidth]{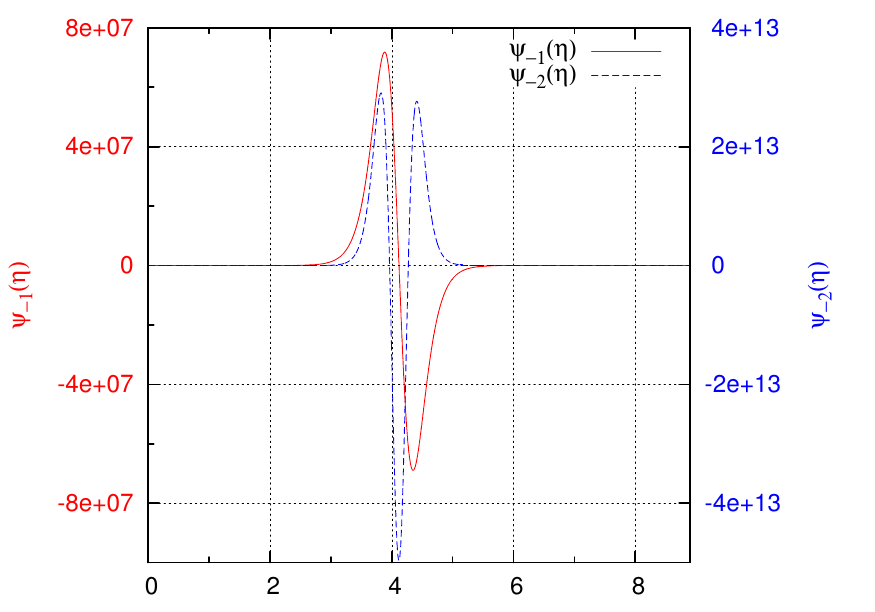}}
\caption{The wave functions of the first two extra negative modes in the case of $B=10$ for the potential (\ref{eq:LW2}).}
\label{fig:modesB10a}
\end{figure}

\begin{figure}
\subfigure[~$\lambda_{-1}$ and $\lambda_{-2}$]{\includegraphics[width=.6\textwidth]{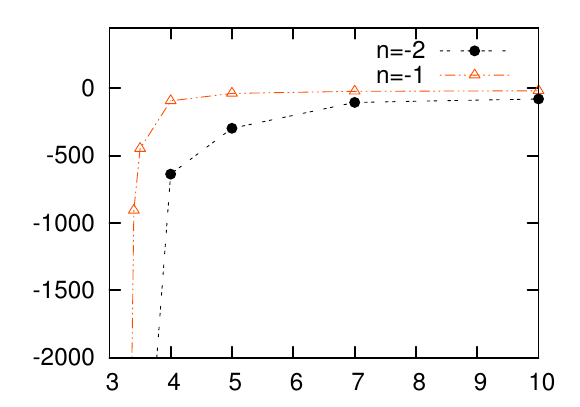}}
\caption{The first two extra negative eigenvalues $\lambda_{-1}$ and $\lambda_{-2}$ of the fluctuation equation (\ref{eq:fe}) versus $B$ for the potential (\ref{eq:LW2}).}
\label{fig:table_a}
\end{figure}

For values of $B>B_0,$ the kinetic pre-factor $Q$ has negative values in some interval and infinitely many additional negative
modes are expected to have support here \cite{Lavrelashvili:1985vn,Lee:2014uza}. For the first time, we will exhibit such modes explicitly here. The wave functions of the first two additional negative modes are shown in Fig.~\ref{fig:modesB10a} for the particular value of $B=10$. As expected, these modes have support in the region over which $Q$ is negative. The two modes shown here have eigenvalues $\lambda_{-1}=-18.22$ and $\lambda_{-2}=-78.45$ respectively. As is intuitively clear, the magnitude of these eigenvalues is inversely proportional to the length of the interval over which $Q$ becomes negative. In Fig.~\ref{fig:table_a} we show how the eigenvalues of the first two additional negative modes evolve as the parameter $B$ in the potential is varied. The eigenvalues already reach $\lambda_{-1}=-907,\, \lambda_{-2}=-9537$ at $B=3.4$ and tend to $-\infty$ as $B$ approaches $B_0$ from above, while for $B<B_0$ these additional negative modes disappear altogether as $Q$ remains positive throughout. Note that over a substantial range of $B$ the magnitude of the eigenvalues of the additional negative modes is vastly larger than those of the tunnelling negative mode and the first couple of positive modes.


\begin{figure}
\subfigure[~$\kappa=0.9$]{\includegraphics[width=.325\textwidth]{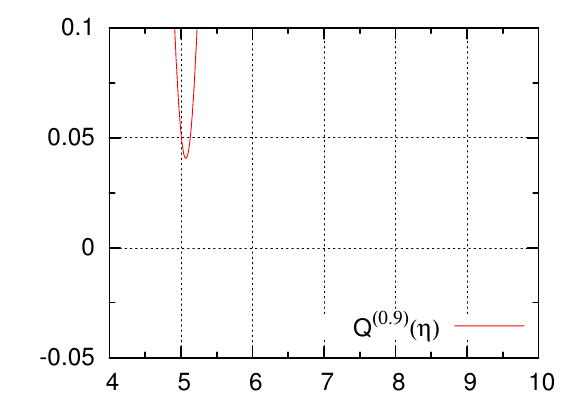}}
\subfigure[~$\kappa=0.7$]{\includegraphics[width=.325\textwidth]{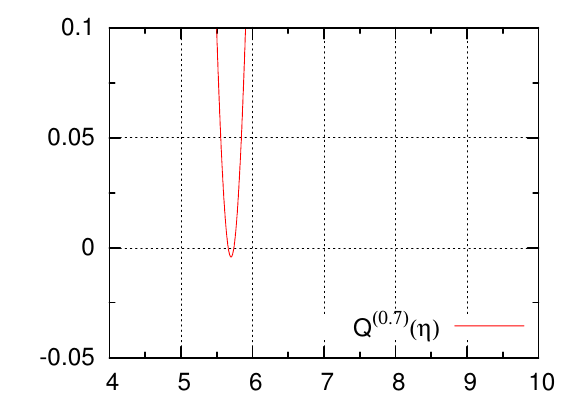}}
\subfigure[~$\kappa=0.5$]{\includegraphics[width=.325\textwidth]{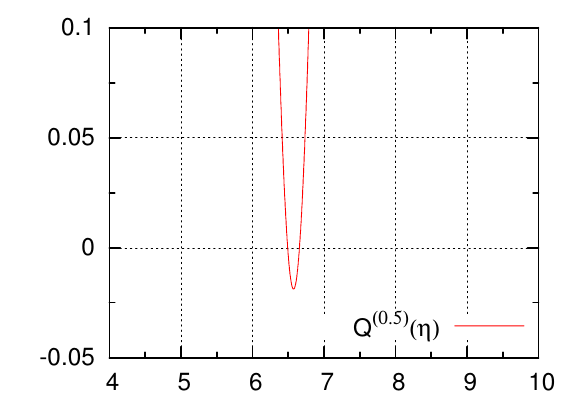}}
\caption{The kinetic pre-factor $Q(\eta)$ (plotted only over the region of greatest interest) for decreasing values of $\kappa$ with fixed parameter $B=3$ for the potential (\ref{eq:LW2}).}
\label{fig:Q-B3}
\end{figure}

\begin{figure}
\includegraphics[width=.495\textwidth]{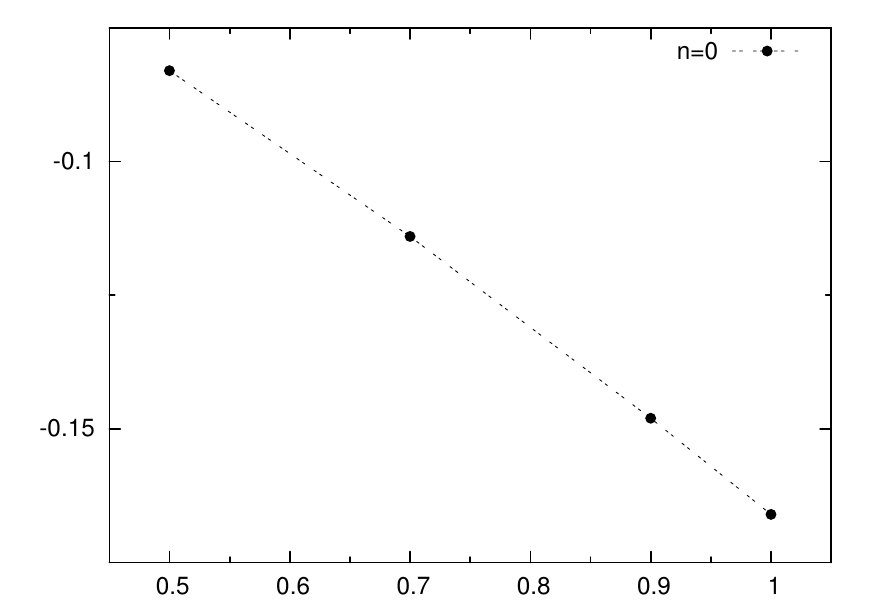}
\caption{Dependence of the eigenvalue of tunnelling negative mode
$\lambda_0$ on different choices of $\kappa$ for $B=3$ for the potential (\ref{eq:LW2}). The eigenvalue evolves smoothly and remains negative.}
\label{fig:B3kappa}
\end{figure}

We also studied the dependence of the above cases on the gravitational strength $\kappa$.
As an example, we fixed $B=3$ and decreased the value of $\kappa$.
Fig.~\ref{fig:Q-B3} shows plots of $Q$ for different values of $\kappa$ when $B=3$.
Fig.~\ref{fig:B3kappa} shows the dependence of $\lambda_0$ with respect to different choices of $\kappa$. Once again, we can see that nothing special happens to the tunnelling negative mode as $Q$ reaches negative values, and in particular the eigenvalue evolves smoothly. As before, we would expect the existence of an additional tower of negative modes wherever $Q$ is negative.



\subsection{Symmetric potential}

\begin{figure}
\subfigure[~$B=2$ and $B=2.5$]{\includegraphics[width=.495\textwidth]{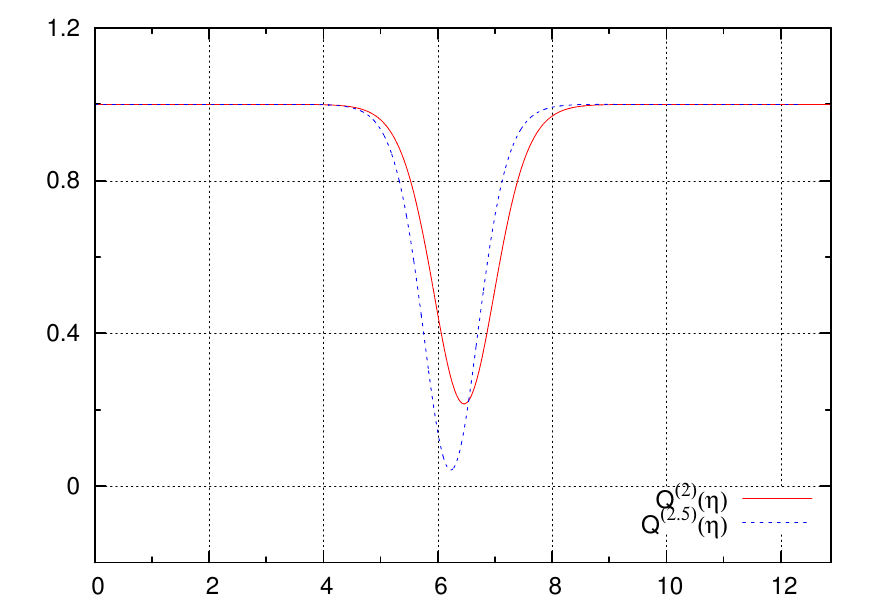}}
\subfigure[~$B=2.75$ and $B=3$]{\includegraphics[width=.495\textwidth]{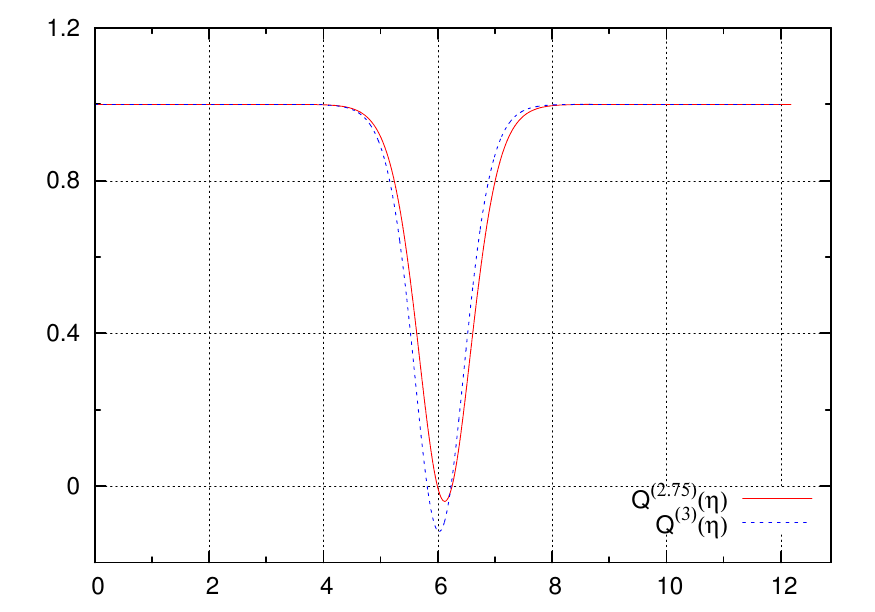}}
\caption{The quantity $Q(\eta)$ for different values of $B$ for
the symmetric potential (\ref{eq:LW3}).}
\label{fig:Qsymm}
\end{figure}

\begin{figure}
\includegraphics[width=.495\textwidth]{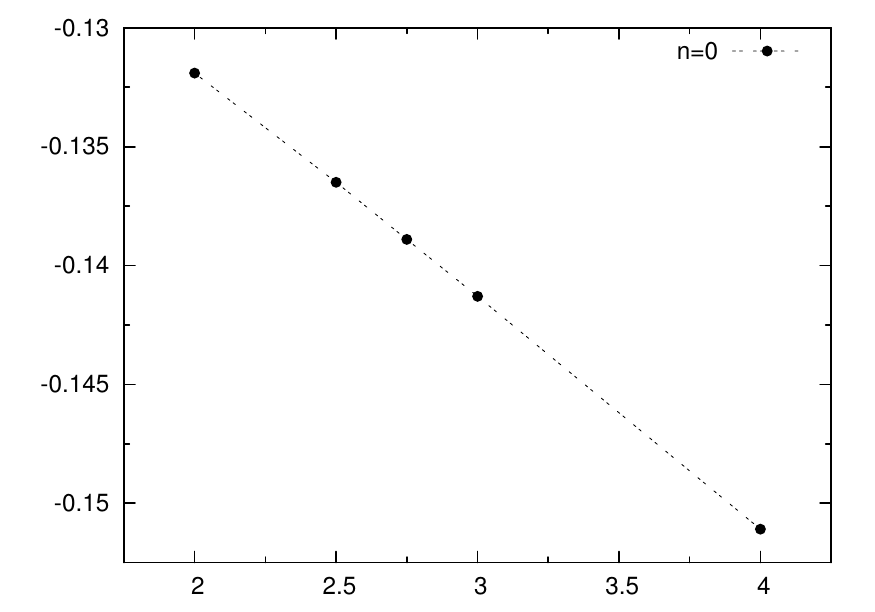}
\caption{Dependence of $\lambda_0$ on different choices of $B$ for the
symmetric potential (\ref{eq:LW3}).}
\label{fig:tablesymm}
\end{figure}

As the third example we consider a symmetric potential. This case is interesting because, in flat space, finite-action Euclidean solutions in a symmetric potential have at most zero modes. When gravity is included, it was first found in \cite{Battarra:2012vu} that the corresponding instantons develop a negative mode. For the numerical investigation, we choose the potential
\be \label{eq:LW3}
V(\varphi) = B (\varphi^2 - \frac{1}{4})^2 + \frac{1}{10}
\ee
and we vary the parameter $B$ in the interval $[2,4]$.
Once again we find a single tunnelling negative mode in all cases, as well as a tower of additional negative modes for parameter values for which $Q$ becomes negative.
Fig.~\ref{fig:Qsymm} shows the kinetic pre-factor $Q$ for selected values of $B$ and
Fig.~\ref{fig:tablesymm} illustrates the dependence of $\lambda_0$ on $B$. The eigenvalue evolves continuously and remains negative as $Q$ first reaches zero and then takes on negative values along an interval of Euclidean time $\eta$ along the instanton.


\section{Discussion}

We investigated tunnelling transitions with gravity with emphasis on
the cases where the factor $Q,$ which sits in front of the kinetic term of fluctuations about the instantons,
is negative in some interval of Euclidean time. We have shown that the perturbation equations remain well-defined when $Q$ crosses zero, and that the eigenvalues of existing mode functions vary continuously as we pass from situations with $Q>0$ everywhere to $Q<0$ somewhere. Thus, in particular, all bounces contain a single tunnelling negative mode in their spectrum of perturbations. We should point out that these results are partly at odds with a recent study of Lee and Weinberg \cite{Lee:2014uza,Lee:2014_PhD} -- we provide a comparison of the two approaches in appendices \ref{appLW} and \ref{appLag}.

As we have discussed, when $Q<0$ an additional infinite tower of negative modes appears,
in analogy with the first description of the problem by LRT \cite{Lavrelashvili:1985vn}.
In the present study we found the first few of these additional negative modes for concrete potentials
numerically and determined their eigenvalues.
We observed that over a substantial range of parameters the magnitude of the eigenvalues of the additional negative modes 
is vastly larger than those of the tunnelling negative mode. 
In our opinion this is another manifestation of the fact that they have a different nature. Furthermore, it is interesting to note that nodal theorems \cite{aq95} are not fulfilled in this case,
namely the number of nodes of the zero energy wave function
does not count  the number of negative energy states in a given potential anymore.

As is known in certain cases, a proper choice of variables (obtained with the help of canonical transformations in the Hamiltonian formalism)
can significantly simplify the description of a system.
In appendix \ref{appCT} we study the influence of canonical transformations on these additional
negative modes and provide arguments that they cannot be removed by a canonical transformation of variables.
The consideration of canonical transformations shows that, even though the pre-factor of the kinetic term cannot be made positive in general,
the classes of instantons for which it is negative can be changed - this suggests that the negativity of the pre-factor is a technical problem,
not related to a fundamental physical cause.
The existence and significance of the additional negative modes remain somewhat mysterious,
and further work will be required to fully elucidate the negative mode problem.

\section*{Acknowledgements}
\par
We are thankful to Lorenzo Battarra for collaboration at the early stages of the present study
and to Chandrashekar Devchand, Marc Henneaux and Valery Rubakov for fruitful discussions. 
M.K.~is supported by the DOE under contract No.~DE-SC0007901. 
G.L.~and J.L.L.~gratefully acknowledge the support of the European Research Council
via the Starting Grant Nr.~256994 ``StringCosmOS''. 
G.L.~acknowledges support from the Shota Rustaveli NSF Grant No.~FR/143/6-350/14 and Swiss NSF SCOPES Grant No.~IZ73Z0-152581.

\appendix

\section{Comparison to the work of Lee and Weinberg} \label{appLW}

It may be useful to compare our results with those of Lee and Weinberg (LW) \cite{Lee:2014uza,Lee:2014_PhD}, who recently performed a related study of the negative mode problem. Their work was performed in the Lagrangian formalism -- in appendix \ref{appLag}, we discuss the Lagrangian approach in more detail. The main difference between our results and those of LW is that there are cases where LW claim that no tunnelling negative mode is present. In particular, for instantons where the wall of the instanton is close to the maximum of $\rho$ (and where the initial and final vacua are nearly degenerate), they find that the tunnelling negative mode is absent. We disagree with these results, as we have explicitly demonstrated the existence of the tunnelling negative mode for such cases, even for the case of exactly degenerate initial and final vacua (a possible explanation of the discrepancy between our results and those of LW is discussed in appendix \ref{appLag}). Moreover, we have provided a proof that solutions to the eigenvalue equation evolve smoothly from cases where $Q$ is everywhere positive to $Q$ going negative somewhere. This analytic result boosts our confidence that our results are correct. We note that LW also confirm the existence of an additional infinite tower of negative modes in regions where $Q$ is negative, though it should be emphasised that in the Lagrangian approach it is the factor $Q_{LRT} = \dot{\rho}^2 -\frac{\kappa}{6} \rho^2 \dot{\varphi}^2$ that is the relevant one, rather than the function $Q = 1-\frac{\kappa}{6} \rho^2 \dot{\varphi}^2$ that appears in the Hamiltonian formalism used here. This makes a difference, as $Q_{LRT}$ becomes negative near $\dot\rho=0$ for \emph{all} bounces (and thus all bounces will have an infinite number of negative modes in the approach of LW),
while the function $Q$ is only negative for certain classes of bounces. We certainly do agree with LW that these additional modes are not properly understood yet, and that their nature and physical significance must be studied further.

\section{Negative mode problem in Lagrangian approach} \label{appLag}

We will briefly describe the derivation of the equations for linear perturbations in the Lagrangian approach.
For the $O(4)-$symmetric situation, Eq.~(\ref{eq:homogen_metric}),
after fixing the gauge $\Psi=0$  and using the constraint equation to eliminate $A$,
one gets the quadratic action \cite{Lavrelashvili:1985vn}
\be
S_\Phi^{(2)} = 2 \pi^2 \int \frac{\rho^3 d\eta}{2 Q_{LRT}} \bigl[ \dot{\rho}^2 \dot{\Phi}^2 - \frac{\kappa \rho^2 \dot{\varphi} V'}{3} \Phi\dot{\Phi}
+(Q_{LRT} V''+\frac{1}{6}\kappa \rho^2 V'^2) \Phi^2 \bigr]  \pkt
\ee
Integrating by parts and using the background equations of motion, one obtains
\be \label{eq:lrt_qa}
S_\Phi^{(2)} = 2 \pi^2 \int \rho^3 d\eta \bigl[ \frac{\dot{\rho}^2}{2 Q_{LRT}}  \dot{\Phi}^2 + \frac{1}{2} U_\Phi \Phi^2 \bigr]
\ee
with the potential
\be
U_\Phi = \frac{\dot{\rho}^2 V''}{Q_{LRT}}  + \frac{\kappa \rho^2 \dot{\rho}^2 V'^2}{3 Q_{LRT}^2}+\frac{\kappa \rho \dot{\rho} \dot{\varphi} V' }{3 Q_{LRT}^2} \pkt
\ee
The exact form of the fluctuation operator depends on the choice of a weight function which can be specified by defining the norm. With the choice
\be
|| \Phi ||^2 \equiv \int d^4 x \  \sqrt{g} \ \Phi^2 = 2 \pi^2 \int d\eta \ \rho(\eta)^3 \ \Phi^2 \kma
\ee
the fluctuation equation diagonalising the quadratic action Eq.~(\ref{eq:lrt_qa}) has the form
\be \label{eq:perturbLRT}
-\frac{1}{\rho^3}\frac{d}{d\eta}(\frac{\rho^3\dot{\rho}^2}{Q_{LRT}}\frac{d\Phi_n}{d\eta}) + U_\Phi [\varphi(\eta),\rho(\eta)] \Phi_n = \lambda_n \Phi_n \kma
\ee
where $\Phi_n$ and $\lambda_n$ are eigenfunctions and eigenvalues of the Dirichlet boundary value problem.
The potential $U_\Phi$ close to $\eta=0$ behaves as
\be
U_\Phi= U_\Phi^{(0)} + U_\Phi^{(2)} \eta^2 + O(\eta^4)
\ee
with
\be
U_\Phi^{(0)}= V''(\varphi_0) -\frac{4}{3}\kappa V(\varphi_0) \kma \ {\rm and} \
U_\Phi^{(2)}=
\frac{5\kappa}{12} V'^2(\varphi_0) + \frac{1}{8}V'(\varphi_0) V'''(\varphi_0) \pkt
\ee
Regular solutions (eigenfunctions) close to $\eta=0$ then admit the asymptotic behaviour
\bea
\Phi &=& D_0 \{ 1 + \frac{1}{8}(V''(\varphi_0)-\lambda) \ \eta^2 +
\frac{1}{576}[ 3 (V''(\varphi_0)-\lambda)^2+ 2\kappa (V''(\varphi_0)-\lambda) V(\varphi_0)  \nn \\
 &+&10 \kappa V'^2(\varphi_0) +3 V'(\varphi_0) V'''(\varphi_0) ] \ \eta^4 + O[\eta^6] \} \kma
\eea
with $D_0$ being a normalisation constant.

This gauge-fixed approach can easily be promoted to the gauge-invariant approach
adopted by Lee and Weinberg (LW) \cite{Lee:2014uza}.
For the gauge-invariant variable
\be
\chi = \dot{\rho} \Phi +\rho \dot{\varphi} \Psi
\ee
one gets a simple quadratic action
\be \label{eq:lag_qa}
S^{(2)}_\chi = 2\pi^2 \int \rho^3 d\eta \left( \frac{1}{2 Q_{LRT}} \dot{\chi}^2 + \frac{1}{2} U_\chi \chi^2 \right) \kma
\ee
with the potential
\be
U_\chi= \frac{V''}{Q_{LRT}}+\frac{\kappa \rho^2 V'^2}{3 Q_{LRT}^2} +\frac{\kappa \rho \dot{\varphi} V'}{3\dot{\rho} Q^2_{LRT}}
+ \frac{2\kappa \dot{\varphi}^2}{3 Q_{LRT}}
-\frac{\kappa \rho \dot{\varphi} V'}{\dot{\rho} Q_{LRT}} -\frac{4 \kappa V}{3 Q_{LRT}} -\frac{\ddot{\rho} \dot{Q}_{LRT}}{\dot{\rho} Q_{LRT}^2} \pkt
\ee
Choosing the norm to be given by
\be
|| \chi ||^2 \equiv \int d^4 x \  \sqrt{g} \ \chi^2 = 2 \pi^2 \int d\eta \ \rho(\eta)^3 \ \chi^2 \kma
\ee
the fluctuation equation takes the form
\be \label{eq:perturbL}
-\frac{1}{\rho^3}\frac{d}{d\eta}(\frac{\rho^3}{Q_{LRT}}\frac{d\chi_n}{d\eta}) + U_\chi [\varphi(\eta),\rho(\eta)] \chi_n = \lambda_n \chi_n \kma
\ee
where $\chi_n$ and $\lambda_n$ are eigenfunctions and eigenvalues of the Dirichlet boundary value problem. For completeness, we will also write out the asymptotic behaviour:
the potential $U_\chi$, close to $\eta=0,$ behaves as
\be
U_\chi= U_\chi^{(0)} + U_\chi^{(2)} \eta^2 + O(\eta^4) \kma
\ee
with
\bea
U_\chi^{(0)}&=& V''(\varphi_0) -\frac{4}{3}\kappa V(\varphi_0) \kma \nn \\
U_\chi^{(2)}&=&\frac{\kappa}{24} V'^2(\varphi_0)
+\frac{\kappa}{3}V(\varphi_0) V''(\varphi_0)
-\frac{2 \kappa^2}{3} V^2(\varphi_0)
+ \frac{1}{8}V'(\varphi_0) V'''(\varphi_0) \pkt
\eea
Regular solutions (eigenfunctions) close to $\eta=0$ behave as
\bea
\chi &=& \chi_0 \{ 1+\frac{1}{8}(V''(\varphi_0)-\frac{4}{3}\kappa V(\varphi_0)-\lambda) \ \eta^2
+\frac{1}{1728} [9 V'(\varphi_0) V'''(\varphi_0)+ 9(V''(\varphi_0)-\lambda)^2 \nn \\
&-& 6\kappa V(\varphi_0) (5 V''(\varphi_0)-9 \lambda ) + 3 \kappa V'^2(\varphi_0)
+8 \kappa^2 V^2(\varphi_0)] \ \eta^4 + O[\eta^6] \} \kma
\eea
with $\chi_0$ a normalisation constant.

We will finish with a simple, but important, remark: the definition of $\chi$ is gauge invariant, but let's pick the typical gauge where one sets $\Psi=0,$ so that we have the relation $\chi = \dot\rho \Phi.$ Now when $\Phi$ describes the lowest eigenmode, i.e.~the negative mode, it has a dependence on $\eta$ in the shape of a typical bell curve, without nodes. However, note that at the middle of the instanton $\dot\rho=0,$ so that $\chi$ will thus {\it necessarily} possess an extra node compared to $\Phi$, even for the lowest-lying eigenfunction! The lowest-lying eigenfunction is peaked near the wall of the instanton, i.e.~near the region where the scalar field $\varphi$ varies the most. If this region is well-separated from $\dot\rho=0,$ one may therefore not see the node, and it may look like $\chi$ is nodeless. However, when the wall is located near $\dot\rho=0,$ the negative mode, expressed in terms of $\chi$, will have one node and will naively look like the first excited mode. In this case, one must be careful not to discard it erroneously. This observation may in part explain the discrepancy between our results and the claims in \cite{Lee:2014uza}.

\section{Canonical Transformations} \label{appCT}

The classical equations of motion of a system can be derived by requiring the stationarity of the action, the integral of the Lagrangian function,
\be
\delta S = 0 \rightarrow \delta \int \left( p q_{,t} - H \right) \mathrm{d}t \,,
\ee
where $(q,p)$ are the canonical coordinate and momentum, while $H$ is the Hamiltonian of the system. The resulting equations of motion are
\be
q_{,t} = \frac{\partial H}{\partial p}, \quad p_{,t} = - \frac{\partial H}{\partial q} \,.
\ee
We could equally well describe this system with different canonically conjugate coordinates $(\tilde{q},\tilde{p})$ and a new Hamiltonian $K,$
\be
\delta \int \left( \tilde{p} \tilde{q}_{,t} - K \right) \mathrm{d}t \,, \quad \tilde{q}_{,t} = \frac{\partial K}{\partial \tilde{p}}, \quad \tilde{p}_{,t} = - \frac{\partial K}{\partial \tilde{q}} \,.
\ee
These two descriptions are equivalent if the two variational principles agree, i.e.~if
\be
p q_{,t} - H = \tilde{p} \tilde{q}_{,t} - K + \frac{\mathrm{d}F}{\mathrm{d} t} , \label{equalityintegrands}
\ee
where it is important to note that one may add an arbitrary total time derivative to the right hand side of the above equation (we are ignoring scaling transformations here, which would have arisen by allowing the right hand side to also be multiplied by an overall factor). This function $F$ may depend on various combinations of the old and new coordinates -- for us, the particular choice
\be
F = q \cdot p  + F_3 (p, \tilde{q}, t),
\ee
will be useful, where $F_3$ denotes an arbitrary function of the old momentum, the new coordinate and of time. With this choice for $F$, Eq. \eqref{equalityintegrands} becomes
\be
-H = \tilde{p} \tilde{q}_{,t} - K + p_{,t} q + \frac{\partial F_3}{\partial t} + \frac{\partial F_3}{\partial p} p_{,t} + \frac{\partial F_3}{\partial \tilde{q}} \tilde{q}_{,t}\,.
\ee
Since we are treating the old and new coordinates and momenta as independent variables, we can see that the equation above is satisfied if we identify
\be
q = - \frac{\partial F_3}{\partial p}, \quad \tilde{p} = - \frac{\partial F_3}{\partial \tilde{q}}, \quad K = H + \frac{\partial F_3}{\partial t}\,.
\ee

Now let us specialise further to the choice
\be
F_3 = -\frac{1}{d} \tilde{q}p - \frac{1}{2dc_{,t}}p^2\,,
\ee
where $c(t),d(t)$ are explicit, a priori arbitrary, functions of time. We have included a time-derivative term in the denominator of the second term such that $c, d$ are dimensionless functions of time. This leads to the canonical transformations
\be
\begin{matrix}  q =  & \frac{1}{d}\tilde{q} + \frac{1}{c_{,t}} \tilde{p} \\ p =  & d\tilde{p}  \end{matrix} \qquad , \qquad \begin{matrix}  \tilde{q} =  &dq - \frac{1}{c_{,t}}p \\ \tilde{p} =  & \frac{1}{d}p  \end{matrix} \qquad , \qquad K = H(\tilde{q},\tilde{p}) - \frac{1}{2} \frac{\partial(d/c_{,t})}{\partial t} \tilde{p}^2\,.
\ee
The important change is in the last term: the momentum-squared part of the Hamiltonian (which is responsible for the kinetic term in the Lagrangian description) gets modified by this transformation.

Now let us look in more detail at the case we are interested in. The Euclidean action $S_E$ was given in Eq. \eqref{qa} (where we will drop the overall factor of $2\pi^2$ arising from the volume of $S^3$). Its Lorentzian equivalent is given by
\be
S_t = \int dt \, {\cal L}_{t} = \int dt \left[ \frac{\rho^3}{2Q} \Phi_{,t}^2 - \frac{\rho^3}{2}U \Phi^2 \right] \,,
\ee
where $Q, U$ are here understood to be given in terms of ordinary time $t=-i\eta,$ e.g. $Q = 1 + \rho^2 \phi_{,t}^2/6.$ With the canonical momentum given by $\Pi = \frac{\rho^3}{Q}\Phi_{,t} ,$ the corresponding Hamiltonian reads
\be
H = \Pi \Phi_{,t} - {\cal L}_t = \frac{Q}{2\rho^3} \Pi^2 + \frac{\rho^3 U}{2} \Phi^2 \,.
\ee
If we now apply the canonical transformation above, with $(\Phi,\Pi) = (q,p) \rightarrow (\tilde{q},\tilde{p}),$ we obtain the new Hamiltonian
\be \label{eq:newHamiltonian}
K = \frac{1}{2} \left[ \frac{d^2 Q}{\rho^3}  + \frac{1}{c_{,t}^2} \rho^3 U - (d/c_{,t})_{,t} \right] \tilde{p}^2  + \frac{ \rho^3 U}{c_{,t}d} \tilde{p}\tilde{q}  + \frac{1}{2} \frac{\rho^3 U}{d^2} \tilde{q}^2\ .
\ee
This theory can equally well be described by a Lagrangian of the form
\be
{\cal \tilde{L}} = \frac{1}{2 \left[ \frac{d^2 Q}{\rho^3}  + \frac{\rho^3}{c_{,t}^2} U - (d/c_{,t})_{,t}\right]} \, \Phi_{,t}^2 + (\cdots) \, \Phi^2 \,,
\ee
where the explicit form of the mass term is not important for our present purposes. Transforming to Euclidean time leads to the Lagrangian
\be
{\cal \tilde{L}}_E = \frac{1}{2 \left[ \frac{d^2 Q}{\rho^3}  - \frac{\rho^3}{\dot{c}^2} U - (d/\dot{c})_{,\eta}\right]} \, \dot{\Phi}^2 - (\cdots) \, \Phi^2 \,,
\ee
where now $Q, U$ take their Euclidean expressions \eqref{eq:Q}, \eqref{eq:Hamiltonian_U}. Note the most important consequence of the canonical transformation: the coefficient in front of the kinetic term has changed. Can it be made positive? Unfortunately, this seems impossible in general: in order to make the kinetic coefficient non-singular, we should take $1/\dot{c} \propto Q$ (with no other zeros than those of $Q$) in order to cancel the divergence in $U$ -- see Eq. \eqref{eq:U}. For some cases of interest, e.g. the potential in Eq. \eqref{eq:LW2} with $B=4,$ $Q^2 U$ is positive everywhere. Then, in the interval where $Q<0,$ the second term in  the kinetic coefficient will make the coefficient even more negative. Hence the sign of the overall coefficient will depend on the last term, $-(d/\dot{c})_{,\eta}.$ At the moments where $Q=0,$ this last term will contribute in proportion to $-d \dot{Q}.$ Keeping in mind that $\dot{Q}$ has different signs at both ends of the $Q<0$ interval, it implies that $d$ must change sign, and hence pass through zero during that interval. But when $d$ passes through zero the $\tilde{p}\tilde{q}$ and $\tilde{q}^2$ terms in Eq. \eqref{eq:newHamiltonian} will be divergent, rendering the new theory singular. There is nevertheless one lesson that one can draw from this discussion: even though this type of canonical transformation cannot make the kinetic coefficient positive and non-singular, it can change the class of instantons for which it is negative. For instance, for instantons where $Q>0$ throughout, one may transform to a theory where the coefficient is negative somewhere. This suggests that the sign of $Q$ might be ``just'' a technical issue, and not something with fundamental physical significance.

There is another set of canonical transformations worth considering. These arise by considering the generating function
\be
F =  F_1 (q, \tilde{q}, t)\,,
\ee
leading to
\be
p = \frac{\partial F_1}{\partial q}, \qquad \tilde{p} = - \frac{\partial F_1}{\partial \tilde{q}}, \qquad K = H + \frac{\partial F_1}{\partial t}\,.
\ee
If we choose $F_1=f(t) q \tilde{q},$ we obtain
\be
q = - \frac{1}{f(t)} \tilde{p}, \qquad p = f(t) \tilde{q}, \qquad K = H + (-\tilde{q}\tilde{p})_{,t} = H\,.
\ee
This transformation interchanges coordinates and momenta. Above, we saw that there are examples, such as the potential \eqref{eq:LW2} with $B=4,$ where the potential is everywhere positive. Hence one may ask whether it is possible to use the transformation just described to exchange a theory with a (partly) negative kinetic term and a positive potential for one with a positive kinetic term and a (partly) negative potential (see also \cite{Gratton:2000fj}).
This would remove the unwanted infinite tower of negative modes. Specialising to the model of interest, we find the resulting Euclidean theory
\be
\tilde{\cal L}_E = \frac{f^2}{2\rho^3 U} \dot\Phi^2 + \frac{Q f^2}{2\rho^3} \Phi^2\,.
\ee
Note that now the kinetic term is indeed manifestly positive when $U>0$ ! However, as it stands the theory is still singular, due to the singularities in $U.$ We can remove those by choosing $f\propto1/Q,$ with no other poles than those of $1/Q.$ This renders the kinetic term regular, but unfortunately leads to a new potential $\propto 1/Q,$ which is divergent when $Q$ passes through zero. Thus, even though we can cure the kinetic term by this procedure, we are in effect shifting the singularity to the potential. Thus, unfortunately it appears that it is impossible to cure the negative $Q$ complications by a canonical transformation.


\end{document}